# Native vs Web Apps: Comparing the Energy Consumption and Performance of Android Apps and their Web Counterparts

Ruben Horn, Abdellah Lahnaoui, Edgardo Reinoso, Sicheng Peng, Vadim Isakov, Tanjina Islam, Ivano Malavolta
Vrije Universiteit Amsterdam, The Netherlands
{r.horn │ a.lahnaoui │ e.j.reinosocampos │ s3.peng │ v2.isakov}@student.vu.nl, {t.islam │ i.malavolta}@vu.nl

*Abstract*—**Context**. **Many Internet content platforms, such as Spotify and YouTube, provide their services via both native and Web apps. Even though those apps provide similar features to the end user, using their native version or Web counterpart might lead to different levels of energy consumption and performance.**
*Goal*. **The goal of this study is to empirically assess the energy consumption and performance of native and Web apps in the context of Internet content platforms on Android.**
*Method*. **We select 10 Internet content platforms across 5 categories. Then, we measure them based on the energy consumption, network traffic volume, CPU load, memory load, and frame time of their native and Web versions; then, we statistically analyze the collected measures and report our results.**
*Results*. **We confirm that native apps consume significantly less energy than their Web counterparts, with large effect size. Web apps use more CPU and memory, with statistically significant difference and large effect size. Therefore, we conclude that native apps tend to require fewer hardware resources than their corresponding Web versions. The network traffic volume exhibits statistically significant difference in favour of native apps, with small effect size. Our results do not allow us to draw any conclusion in terms of frame time.**
*Conclusions*. **Based on our results, we advise users to access Internet contents using native apps over Web apps, when possible. Also, the results of this study motivate further research on the optimization of the usage of runtime resources of mobile Web apps and Android browsers.**

## I. INTRODUCTION

The market share of mobile devices has grown explosively in recent years. Surpassing desktop computers in 2016 and accounting for around 60% of devices as of 2022 [1]. This makes it the primary target for Internet content platforms, which includes any provider that offers information or services over the Internet. Many Internet content platforms, such as Spotify and YouTube, provide their services via both native and Web apps. Native apps are self-contained software packages that are installed on the target device and usually bundled with all static resources, while Web apps are composed of HTML documents, CSS style sheets, and JavaScript for client-side computation. Web apps are accessed through a Web browser and their resources are downloaded every time the Web app is opened unless cached. Dynamic content is loaded at runtime, typically over HTTP(S) from a remote endpoint. Content platforms have the incentive to provide their services via both native and Web apps, as seen in Fig. 1, with full or nearly full feature parity in order to capture as many users as possible. Even though those apps provide similar features to the end user, using their native version or Web counterpart might lead to different levels of energy consumption and performance.

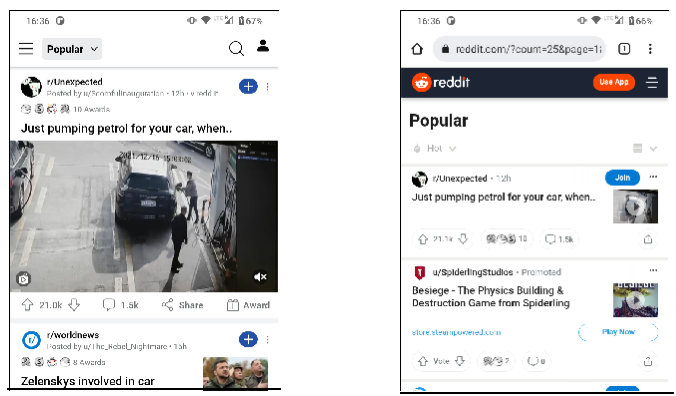

Fig. 1: Reddit native Android (left) vs Web app (right)

Users can be motivated to use either one due to various factors, such as their environmental awareness, improved battery life, and perceived responsiveness contributing to better usability. One should not however assume that this will be a universal decision since noteworthy popular platforms vary greatly in the type and method of presenting their content. For example, YouTube and TikTok offer almost exclusively video, while other social media platforms such as Facebook, Twitter, and Reddit consist of text, images, and video combined. A very significant part of using such platforms consists of consuming the content that they provide, which is possible either through a native app or the corresponding mobile Web app. The observation that video content requires more energy and processing than just audio playback is trivial. Users can infer this from the difference in battery life for the different activities. Some manufacturers, like Apple, also report this in the expected battery life for different activities in the device specifications [2]. Gauging the extent of the difference in performance and energy consumption between native apps and their Web app counterparts is a non-trivial research avenue. The answer to this could be useful in deciding which version of any given platform a user should consider.

The **goal** of this study is to compare the energy consumption and performance of native Android apps and their Web counterparts. To this aim, we select 10 Internet content platforms across 5 categories. Then, we measure them based on the energy consumption, network traffic volume, CPU load, memory load, and frame time of their native and Web versions; then, we statistically analyze the collected measures and report our results.

Based on our results, we advise users to access Internet content using native apps over Web apps, when possible to optimize the battery life of their devices. The maintenance of two different versions of the same app (perhaps even more in the case of multiple native apps for different platforms) requires additional resources from the provider. Therefore, the results might be of interest to the developers to decide which type of app is best suited for their particular platform. Focusing only on one version may allow them to lower development costs. Additionally, the findings of this study will motivate further research on the optimization of the usage of runtime resources of mobile Web apps and Android browsers.

## II. EXPERIMENT DEFINITION

The goal of this study is to *analyze app types for the purpose of evaluation with respect to their energy consumption and performance from the point of view of a user in the context of Internet content platforms on Android*.

From the goal, we derive the following two research questions which focus on energy and performance respectively:

> **RQ1** *How does energy consumption vary between native and Web versions of the same app?*

Between native and Web, technical differences such as the media codecs, content (pre-)fetching and caching strategy or user interface behavior can impact the energy consumption and performance of mobile apps. We suspect that in practice, there may be a statistically significant difference in energy consumption and performance. In order to answer the question quantitatively, the energy consumption of the apps is measured in Joules (J) over a certain period of time while interacting with the content in a typical fashion. The typical user interaction may vary between apps, depending on how much input is required to consume the content. To simulate continuous consumption of content, we provide a custom input sequence per content platform, for example, continuous scrolling to load new items in a news feed or no input when watching a video. Since the quality of the content may differ between native and Web versions (compression, available resolution, protocol overhead), we also measure the network traffic volume.

> **RQ2** *How does performance vary between native and Web versions of the same app?*

Native Android apps are predominantly written in Java, while Web apps run within the HTML and JavaScript engine of the browser app. This could result in computational overhead, however the performance of user interface components and their use by Web developers could even be more optimized compared to native apps. For this research question, we consider the utilization of the hardware resources (CPU, memory) and the achieved refresh rate of the user interface as well as the network traffic volume. CPU utilization is measured as percentages of the respective maximum device capability and memory utilization in kilobytes (kB). The refresh rate is the time between two consecutive frames being rendered (ns) and the network traffic volume is the total amount of bytes (B) sent and received by the device. A lower frame rate could be an indication of insufficient hardware resource utilization or availability, while a higher network traffic volume at similar hardware utilization may indicate more efficient data processing. Both CPU and memory utilization are measured because some, but not all, content is strictly linear (video).

## III. EXPERIMENT PLANNING

### A. Subjects Selection

For the selection of subjects in this research, we start by obtaining a list of the top 2000 most visited fully qualified domain names from the Tranco list [3], as well as a list of the top 2000 most downloaded native apps from the Google Play Store from a Kaggle dataset [4]. Next, we create a pairwise matching between the two lists using the domain name without the top-level domain and the app name based on exact lowercase string comparison per word and filter out all elements for which no match could be made. This is necessary because sometimes the app name contains additional words. Matching on the package name identifier did not yield correct results, because some companies changed their app name and domain used on the Web as part of a rebranding, but the package name cannot be changed. From this, we obtain a list of 170 domains without duplicates. The category of each platform is derived from the app category in the Google Play store and described in Table I. Other categories like utilities or games are excluded because they are not suitable for this study; due to the functional diversity of the apps within the respective category, the lack of dynamically loaded content from a remote source, or the lack of a Web app version. For each of our categories in Table I we randomly sample two items. This is done by generating a random permutation of the entire list and iterating over it. For each item, we do the following steps until we have selected 10 subjects:

1) Verify that comparable native and Web apps exist for Android
2) Determine the category according to Table I
3) Select the current item, if less than two items have been selected for the category of the current item (see Table I)

TABLE I: Selected content platform categories

| Category | Description |
| --- | --- |
| News | Online newspapers, magazines as well as specialized offerings such as in weather, finance, and sport |
| Social media | User driven platforms including messaging, (micro-)blogging and image boards |
| E-Commerce | Online retail, review, and trading platforms |
| Audio streaming | Playback of recordings or live streams |
| Video streaming | Playback of recordings or live streams |

While the number of investigated apps is rather small, having more than one app per category may help to support the generalizability of any findings. We only consider native Android apps and Web apps running in the Android version of the Google Chrome browser to limit the scope of the study. Many default app settings vary between native and Web versions, such as automatically playing video previews and default resolution. To simplify the experiment, we consider the default app settings assuming that they are not changed by the majority of users and are consciously set by the developers. Further, only apps that can be used with a free account without providing a phone number and without regional restrictions are considered to simplify the experiment. False positive matches using similar names are manually removed. Using this method, we obtain the following study subjects listed in Table II.

TABLE II: Subjects

| Native app | Web app | Category |
|---|---|---|
| ESPN | espn.com | News |
| The Weather Channel | weather.com | News |
| LinkedIn | linkedin.com | Social media |
| Pinterest | pinterest.com | Social media |
| Coupang | coupang.com | E-Commerce |
| Shopee | shopee.tw | E-Commerce |
| SoundCloud | soundcloud.com | Audio streaming |
| Spotify | spotify.com | Audio streaming |
| Twitch | twitch.tv | Video streaming |
| YouTube | youtube.com | Video streaming |

### B. Experimental Variables

In the following paragraphs, we identify the independent and dependent variables for our two research questions. The dependent variables are shown in Table III.

TABLE III: Dependent variables

| Variable | Description | RQ |
|---|---|---|
| Energy consumption ($e$) | Energy consumption is measured in Joules (J) as the energy consumed by the mobile device during the experiment run | RQ1 |
| Network traffic ($n$) | Amount of data in Bytes (B) sent and received by the mobile device during the experiment run | RQ2 |
| CPU load ($c$) | Mean relative (%) device CPU utilization across all cores | RQ2 |
| Memory load ($m$) | Mean (kB) device memory utilization | RQ2 |
| Frame time ($f$) | Median time in nanoseconds (ns) between two successive frames (We use median as an aggregation measure, since we expect extreme outliers due to apps blocking on the main thread during certain operations) | RQ2 |

For our first research question concerning energy consumption, we only have one independent, nominal variable *APP TYPE*. The two possible treatments for the factor corresponding to this variable are 'native' and 'Web'. This determines if the user should interact with the native or Web app for the respective content platform. Thus, each run of a fixed length of 3 minutes selects the concrete app and version for simulated interaction based on the treatment for this factor. The single dependent, continuous variable for the first research question is the energy consumption ($e$) of the device.

For our second research question, we have the same independent variable *APP TYPE* and possible corresponding values as for the first research question. Here multiple dependent variables indicate the performance. CPU load ($c$) and frame time ($f$) are continuous while memory load ($m$) and network traffic volume ($n$) are discrete. These variables are captured using different plugins for the Android Runner experiment framework [5], namely the *batterystats*, *Android* and *frametimes* plugins provided by the framework itself. Network traffic volume is measured in bytes using a custom script that computes the difference of the overall traffic volume measured by the operating system using the *dumpsys* utility [6].

### C. Experimental Hypotheses

Let $\mu_{d_t}$ denote the mean of the sample for a given dependent variable ($d$) and *APP TYPE* ($t$). For our first research question, the null hypothesis is that the mean energy consumption for any native app is equal to its Web app counterpart. The alternative hypothesis is consequently that there is a difference between native and Web versions of the same app in terms of energy consumption. Since we do not know if native apps could consume more or less energy than Web apps, we use a two-sided statistical test.

$$\begin{aligned} H_0 &: \mu_{e_{\text{native}}} = \mu_{e_{\text{Web}}} \\ H_a &: \mu_{e_{\text{native}}} \neq \mu_{e_{\text{Web}}} \end{aligned} \quad (1)$$

For our second research question, we investigate the performance indicated by network traffic volume ($n$), CPU load ($c$), memory load ($m$), and frame time ($f$). The null-hypothesis is that there is no difference in means for any of these four dependent variables between native and Web versions of a subject. The alternative hypothesis states that there is at least one variable that is significantly different between native and Web versions. Once again, we do not make any prior assumption about the sign of the difference between native and Web, so a two-sided statistical test is required.

$$\begin{aligned} H_0 &: \mu_{d_{\text{native}}} = \mu_{d_{\text{Web}}} \quad \forall d \in \{n, c, m, f\} \\ H_a &: \mu_{d_{\text{native}}} \neq \mu_{d_{\text{Web}}} \quad \exists d \in \{n, c, m, f\} \end{aligned} \quad (2)$$

### D. Experiment Design

Our experiment only has a single factor, leading to a simple design. We consider 5 different categories, as described in Table I. Furthermore, we have 2 different app types: native and Web apps. Last, for each category, two subjects will be evaluated, for example, YouTube and Twitch in the category of video streaming platforms. As a result, the total number of trials for the experiment is 20. Moreover, the total run duration is chosen to be 3 minutes (180 seconds). This is a significant length that allows for capturing meaningful data for our measurements. In order to take into account the possible fluctuations of the collected energy measures and to reach higher statistical power [7], the number of repetitions for each subject is 25 times, resulting in a total of 500 runs over all subjects. This should help to average out our results, thereby reducing the impact of small variations between runs. The cooldown process ensures that energy consumption, power state, as well as the temperature of the device to return to ambient levels after each run. In order to achieve this, we pause the experiment execution for 30 seconds after each run.

The Android Runner framework furthermore adds a framework overhead of 60 seconds to each run. Finally, the total duration for the execution of the experiment comes out to 37.5 hours, considering all the aforementioned factors.

*E. Data Analysis*

Since zero or negative values are not possible for the selected metrics, any run that either contains such values or is missing any values indicates some failure in the execution. These runs are thus not included in the analysis to avoid inaccurate results. For each subject, we ensure that we have the same number of data points for the native and Web version. We discard additional runs by random selection. For all tests, we use a standard likelihood threshold $\alpha = 0.05$. In the context of our experiment, the population is the set of all pairs of corresponding native and Web apps available on Google Play and the Web. After obtaining the dependent variables for each run from the measurements, they are quantitatively analyzed.

*a) Data description and exploration:* Initial insight is obtained by visualizing the data per dependent variable and treatment using box-jitter mixed plots.

*b) Testing for normality:* Using the Shapiro-Wilk test, we determine if the data for each dependent variable follows a normal distribution over all values for *CATEGORY*. Visual inspection is performed by means of density and quantile-quantile (QQ) plots.

*c) Given normality:* We investigate the difference between native and Web versions for all categories by performing a paired t-test. Since we do not make any prior assumption about the sign of the residual, we perform a two-tailed test. We quantify the effect size using Cohen's d measure if a significant difference can be found and $H_0$ from hypotheses set 1 can be rejected. The obtained effect size is interpreted according to the suggestion of [8].

*d) Not given normality:* We use non-parametric tests to investigate the difference between native and Web versions. First, we do not consider the category and apply the Wilcoxon signed-rank test and, if a statistically significant difference is found. In this case, we reject $H_0$ from hypotheses set 2. To quantify the effect, we use Cliff's delta. Our interpretation is informed by [9].

## IV. EXPERIMENT EXECUTION

A replication package of the code used for the experiment is made available through an anonymous repository [10].

*A. Preparation*

The first step consists of downloading the list of native apps that were selected as subject of study in Section III. For this, we download the corresponding app from the Google Play Store and extract the APK file using the Android Debug Bridge (ADB) [11]. For convenience, a simple bash script is used to install or uninstall all the APK files prior to running the experiment. This guarantees that the correct app version (Native app versions are available in the replication package: see Section IV) is used and saves time during the experiments since apps are not installed and uninstalled between runs. Furthermore, by first removing all subjects from the device, we guarantee that they will use their default configuration during the experiment. The Web apps are run in the Google Chrome browser app. This Web browser is chosen because it has the highest market share on mobile devices [12]. Some Internet content platforms require the user to authenticate in order to engage with their content. Since this is a hard and time-consuming process to automate, and it may further show dynamic user interface elements, part of the initial setup consists of the manual user authentication for native and Web apps. At this point, the experiment is prepared to be setup for the following steps in the execution.

*B. Setup*

Next, we describe our experiment setup, which is visualized in Fig. 2. The experiment is carried out using the experiment framework Android Runner [5] with a Nokia 6.2 (model TA-1198) running Android (Go edition) 10. This entry level smartphone from 2019 is equipped with a 1.8 GHz octa-core Snapdragon 636 CPU, 3 GB memory, and a 3500 mAh battery [13]. The android device (hereafter just device) is connected through USB 3 to a computer with a GNU/Linux 5 system running the experiment framework. This computer is connected to 230V wall power and charges the device. Since the energy consumption of the device is estimated using the hardware activity and power profile of the device, the battery is not drained during the experiments. This also prevents the device from switching on any power saving measures that could impact the experiment or from unexpectedly shutting down. All measurements are taken on the device itself using the built-in debugging tools and transmitted to the computer through the ADB and the *systrace* utility [14]. Using ADB, the device state is configured and input is simulated during the experiment. These features are provided by the Android Runner framework [5], which orchestrates the experiment following the corresponding configuration file. This configuration further specifies scripts and plugins to be run before and after each step in the experiment. It is not feasible to completely mock the remote resources for all subjects in our experiment. Thus, they are accessed regularly over the Internet using Wi-Fi. During the execution of the experiment, we limited fluctuations of the Wi-Fi network by having only one device connected to it, always at the same distance from the Wi-Fi router. Before the experiment, the connection speed is measured to be 100Mbps using *speedtest* [15].

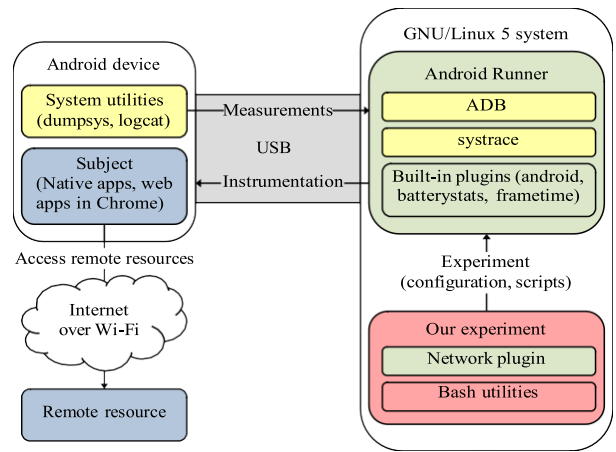

Fig. 2: Experiment setup (yellow indicates Android utilities, blue the subjects, green the components of the Android Runner framework, and red our experiment.)

Using built-in hooks of the Android Runner framework, we ensure that the device is in a pre-defined state for each run. At the beginning of the experiment, a custom script resets the state of the device, which includes the following steps: (1) Unlock the device, (2) Enable 'Stay awake', (3) Stop all running apps, (4) Enable airplane mode but keep Wi-Fi enabled, (5) Dim the screen and mute media playback.

Before each new run, the app of the previous run is shut down using ADB. Since some selected applications require authentication and take measures against repeated automated logins, we log in using a test account before the experiment and do not clear the app data between runs. However, after each run, we automatically selectively clean the browser session including tabs and all cached data except cookies required to persist the authentication. For each run of a native app, we clear the corresponding app cache.

### C. Measurement

Android Runner is used to start each experiment. All parameters for measuring the dependent variables are specified in two configuration files, *config_native.json* and *config_web.json*. These measurements are collected by the built-in android, batterystats, and frametimes plugins as well as our own custom network plugin.

TABLE IV: Tools and rates for the measurements

| Measurement | Rate | Plugin | *dumpsys* utility |
|---|---|---|---|
| CPU load | 10 Hz | Android | *cpuinfo* |
| Memory load | 10 Hz | Android | *meminfo* |
| Energy consumption | n/a | Batterystats | *batterystats* |
| Frame time | 10 Hz | Frametimes | *gfxinfo* |
| Network traffic volume | n/a | Network | *netstats* |

TABLE V: Automated usage scenarios for selected platforms

| Usage scenario (looped) | |
|---|---|
| ESPN | Open news article, scroll down, continue with next article |
| The Weather Channel | Check hourly forecast, check 10-day forecasts, check radar |
| LinkedIn | Scroll personal feed, scroll jobs |
| Pinterest | Scroll posts, open post, go back |
| Coupang | Open category, scroll products, open product page, check comments |
| Shopee | Open category, scroll products, open product page, check comments |
| SoundCloud | Listen to promoted song |
| Spotify | Search for a playlist, listen to playlist |
| Twitch | Search for channel, watch channel |
| YouTube | Search for video, watch video |

Table IV shows how each plugin collects the data using native android utilities. The native/Web apps are then launched and follow a set of instructions outlined in each app's *interaction.py* file. These instructions are a combination of *tap*, *swipe*, and *write-text* commands in order to replicate the usage scenarios mentioned in Table V. Since unpredictable interstitial advertisements can disturb the interaction, we use dns.adguard.com for DNS, which prevents advertisements from being shown without impacting the functionality of the apps. Additionally, there is a 30-second interval between each run to ensure they are properly stopped with no lingering effects.

### D. Analysis

R studio [16] version 2022.2 along with R [17] version 4.1 are used for the statistical tests and analysis including the *effsize* package [18]. The results are visualized using the package *tidyverse* [19] and *qqplotr* [20].

## V. RESULTS

Before analyzing the results obtained from the experiment, all invalid data must be removed. For a considerable portion of runs, the network traffic volume measures 0 bytes. This may be caused by the statistics not being updated before being read by the profiler. Since this metric is completely independent of all others, we simply discard all of its invalid values in the analysis, leaving a reduced amount of data point pairs per subject for this metric. For YouTube, 24 pairs could be used, for Twitch 23. ESPN and Shopee have 19 valid pairs. For Spotify, The Weather Channel and SoundCloud it is only 13, 12 and 11 pairs respectively. For the remaining subjects (Coupang, LinkedIn, Pinterest), no invalid data needed to be removed. Due to the resulting uneven distribution of subjects for this metric, the corresponding measurements may be skewed.

The descriptive statistics for all metrics are shown in Table VI. For the initial exploratory analysis of the experiment data, we visualize the measurements for all dependent variables using box plots with a jitter overlay of the individual data points and density plots in Fig. 3, Fig. 4, Fig. 5, Fig. 6 and Fig. 7.

Using these figures and the corresponding legend in Table VIII, we can already make some interesting observations about the data we have collected.

*a) Energy consumption:* With respect to energy consumption (see Fig. 3), we observe that the native versions of our subjects have a noticeably lower mean and a clearly non-normal distribution, with two distinct peaks compared to the Web versions which have less distinct peaks. While the Web version of Twitch seems to consistently consume the most energy, it is situated in the lower peak within the distribution of the native versions. By contrast, social media and e-commerce apps (which are situated close to the mean consumption of Web apps) can be found towards the top of the distribution of energy consumption for native apps.

*b) Network traffic volume:* Looking at the network traffic volume (see Fig. 4), we observe a slightly lower mean and smaller standard deviation for Web apps. This is intuitively plausible because developers are incentivized to reduce the bandwidth requirements for Web apps, as they often have to be re-downloaded on startup. Appearing noticeably bimodal, the distribution is more spread out for native apps. The high network traffic volume for social networks may be explained by higher-resolution previews and automatic video playback.

TABLE VI: Descriptive statistics for the dependent variables described in Table III of the analyzed runs (*Invalid runs excluded)

| APP TYPE | native | | | | | web | | | | |
|---|---|---|---|---|---|---|---|---|---|---|
| Variable | e | n* | c | m | f | e | n* | c | m | f |
| Mean | 371.1275 | 39422545 | 24.22059 | 238682.8 | 12205412 | 567.4365 | 21956270 | 35.89417 | 1756048 | 13691591 |
| Standard deviation | 63.65198 | 42210598 | 8.961855 | 83187.25 | 7559592 | 65.85269 | 37688979 | 5.890589 | 139139 | 9434142 |
| Minimum | 261.1512 | 473403 | 4.84 | 82531.12 | 6387138 | 435.402 | 929802 | 22.57009 | 1542501 | 4955741 |
| 25% quantile | 313.848 | 2368311 | 17.6412 | 174129.6 | 8823069 | 517.8576 | 4249584 | 32.51875 | 1661566 | 7368804 |
| Median | 359.9316 | 23588223 | 22.77656 | 226458.9 | 9610110 | 559.6747 | 8828234 | 35.9775 | 1697927 | 9386908 |
| 75% quantile | 433.115 | 82428927 | 30.38618 | 289301.5 | 12335778 | 607.3942 | 23655925 | 38.5815 | 1861976 | 15314469 |
| Maximum | 524.8303 | 161970231 | 53.39976 | 444975 | 47830370 | 869.6102 | 349446077 | 53.78788 | 2198063 | 42525146 |

*c) Mean CPU load:* The distribution of the mean CPU load of native apps shown in Fig. 5 ranges from roughly 10% to roughly 50% and looks like a mix between two highly overlapping bell shapes. For Web apps the mean is at slightly above 35% and the distribution has three narrow peaks. The news apps form a peak at the lower end, perhaps likely due to the fact that not much processing is required once the initial content is loaded and displayed. Audio and video streaming exhibit another peak above the mean, most likely due to the continuous decoding of the incoming media.

*d) Mean memory load:* The stark difference in the mean memory load of roughly 1.52 GB between native and Web apps is observable in Fig. 6. This is similar to the findings of [21] and [22]. A possible explanation for the memory overhead could be the fact that Web apps run on top of the Google Chrome browser app. Since the memory footprint of the browser contributes to the measurements, it is plausible that the measurements for the Web apps are higher than those of the native versions. For other browsers, which have a higher memory load, this effect may be even stronger as indicated by [23, Figure 3].

*e) Median frame time:* Looking at the density plot in Fig. 7, there does not seem to be a large difference between native and Web apps with respect to this independent variable. The only consistent outliers are the Twitch Web app and both the native and Web app for SoundCloud. However, falling still below 50 ms, this is not sufficient evidence for degraded performance that would be noticeable by the user.

*A. Testing for normality*

As already indicated by the density plots in Fig. 3, Fig. 4, Fig. 5, Fig. 6 and Fig. 7, the QQ plots available in the replication package (see Section IV) show that for all dependent variables, the data for native and Web apps is not normally distributed. This is further confirmed by the Shapiro-Wilk tests in Table VII which shows non-normal distribution for all dependent variables for both native and Web. Thus, we use non-parametric tests for further analysis.

*B. Hypothesis testing*

The distributions of the energy consumption data points is not normal for both native and Web, thus, we use the

TABLE VII: Results for the Shapiro-Wilk test with $\alpha = 0.05$

| Dependent variable | APP TYPE | p-value | normal |
|---|---|---|---|
| Energy consumption | native | $1.359 \times 10^{-10}$ | no |
| | web | $3.557 \times 10^{-8}$ | no |
| Network traffic | native | $1.047 \times 10^{-14}$ | no |
| | web | $3.373 \times 10^{-24}$ | no |
| Mean CPU load | native | $3.16 \times 10^{-5}$ | no |
| | web | $8.012 \times 10^{-6}$ | no |
| Mean memory usage | native | $3.177 \times 10^{-9}$ | no |
| | web | $7.228 \times 10^{-15}$ | no |
| Median frame time | native | $9.159 \times 10^{-24}$ | no |
| | web | $1.084 \times 10^{-19}$ | no |

Wilcoxon signed-rank test to determine if the two populations are different. Using, $\alpha = 0.05$ we reject the null-hypothesis in (1) that they are the same based on the p-value for the variable *e* in Table IX, which is significantly smaller than $\frac{\alpha}{2}$

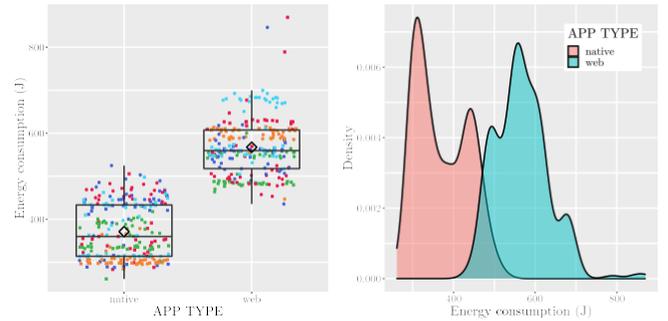

Fig. 3: Box and density plots for energy Consumption

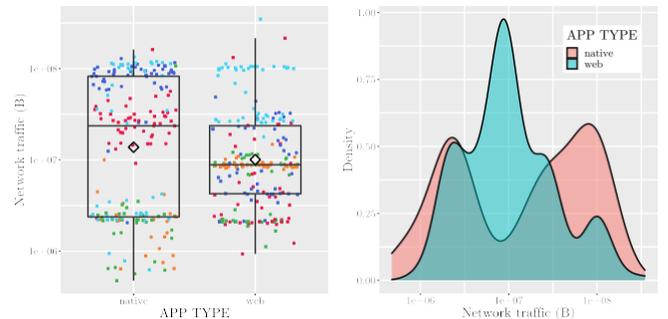

Fig. 4: Box and density plots for network traffic volume using logarithmic scale

For the non-normally distributed dependent variables associated with performance, we once again use Wilcoxon signed-rank tests with $\alpha = 0.05$. For all of them except *f*, the p-value is in Table IX smaller than $\frac{\alpha}{2}$, confirming a difference between the samples. Thus, we can reject the null-hypothesis in (2) that the populations are not different.

TABLE VIII: Legend for data points in jitter plots

| ESPN | The Weather Channel | LinkedIn | Pinterest | Coupang | Shopee | SoundCloud | Spotify | Twitch | YouTube |
|------|---------------------|----------|-----------|---------|--------|------------|---------|--------|---------|
| ● | ■ | ● | ■ | ● | ■ | ● | ■ | ● | ■ |

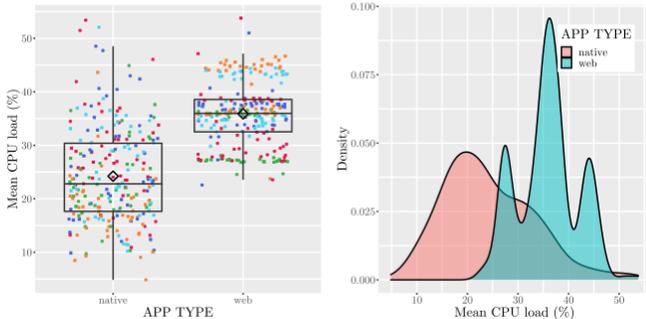

Fig. 5: Box and density plots for mean CPU load

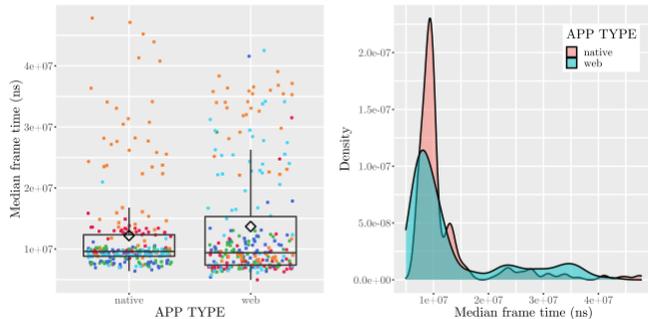

Fig. 7: Box and density plots for median frame time

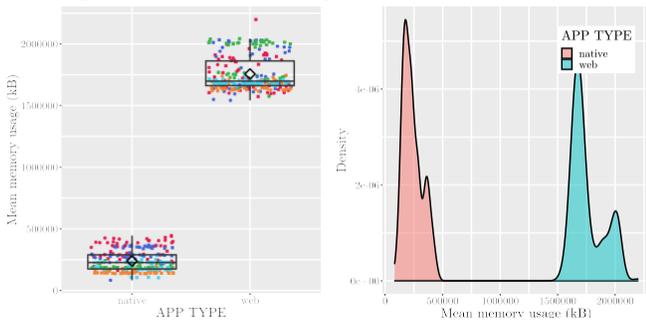

Fig. 6: Box and density plots for mean memory load

### C. Interpretation of effect size

Since we perform exclusively non-parametric tests, we use Cliff's delta in order to quantify any statistically significant difference between native and Web apps as $d$. A 'negligible' difference is indicated by $|d| < 0.147$, a 'small' difference by $0.147 \leq |d| < 0.33$, a 'medium' difference by $0.33 \leq |d| < 0.474$, and a 'large' difference by $0.474 \leq |d|$. The sign indicates the direction of the difference where negative means that the values for native are lower than those for Web and positive indicates the opposite. This interpretation is proposed by [9]. The results of these tests are also shown in Table IX.

TABLE IX: Wilcoxon signed-rank tests for the dependent variables (where $e$ = energy consumption, $n$ = network traffic volume, $c$ = CPU load, $m$ = memory load and $f$ = frame time) with $\alpha = 0.05$ to determine the differences between native and Web with Cliff's delta to gauge the effect size (Interpretation following Romano et al. [9])

| Variable | p-value | Cliff's delta | Effect size | RQ |
|----------|---------|---------------|-------------|-----|
| $e$ | $1.335 \times 10^{-42}$ | -0.98384 | large | RQ1 |
| $n$ | $5.948 \times 10^{-9}$ | 0.1642591 | small | RQ2 |
| $c$ | $4.435 \times 10^{-29}$ | -0.72272 | large | RQ2 |
| $m$ | $9.31 \times 10^{-43}$ | -1 | large | RQ2 |
| $f$ | $5.293 \times 10^{-1}$ | n/a | n/a | RQ2 |

Using the value obtained for Cliff's delta and the interpretation described above, we identify a large difference between native and Web in regard to energy consumption.

For the network traffic volume ($n$) a logarithmic scale is used in Fig. 4 to better visualize the difference. While the median values for native and Web differ by an order of magnitude, the overall range of values in both samples is between 3 and 4 orders of magnitude and the values of the medians are rather large, sitting at either side of $10^7$, so the difference is relatively small. This is confirmed using Cliff's delta, which indicates a small difference. For the CPU utilization ($c$) and memory utilization ($m$) the effect is determined to be large using this measure. The latter even has no overlap in the density plot in Fig. 6. Since the maximum absolute value of Cliff's delta is 1.0, the effect is too extreme to be meaningfully quantified by this metric. Because there is clearly a difference in means between native and Web, we also compute Cohen's d for memory utilization, which results in a value with a magnitude of 13.23714. This also indicates a large effect size according to [8]. For the frame time ($f$), we do not compute a value for Cliff's delta, since there is no statistically significant difference based on its $p$-value.

The shapes of all distributions are rather different, thus, we can not make any assumptions about the difference of the population means based on the results of the non-parametric tests if the distributions overlap to a significant degree, such as in the case of network traffic volume. Nonetheless, we find a (far) more than negligible difference for the other dependent variable. Turning again to the box plots from above, we find a large effect size in those where the interquartile ranges do not overlap. Those are Fig. 3, Fig. 5 and Fig. 6. This intuitively confirms the effect sizes prescribed to them.

## VI. DISCUSSION

Based on the $p$-values of the Wilcoxon signed-rank tests and corresponding effect sizes obtained using Cliff's delta from the previous section, we proceed to answer our research questions.

For **RQ1**, we conclude that native apps consume significantly less energy than their Web counterparts, with a large effect size.

We thus recommend end users access Internet content using native apps over Web apps to optimize the battery life of their Android devices. However, since we did not consider app background activity such as (location) tracking and push notifications, which are mostly exclusive to native apps, exceptions to this rule may exist. Since there is a statistically significant difference in the energy consumption of native and Web apps, this may only occur in extreme cases.

For **RQ2**, the $p$-values are significantly smaller than the $\alpha$ threshold for most dependent variables indicative of performance (network traffic volume, CPU utilization, and memory utilization). Web apps use more CPU and memory, with a statistically significant difference and large effect size. Therefore, we conclude that native apps tend to require fewer hardware resources than their corresponding Web versions. The network traffic volume exhibits a statistically significant difference in favour of native apps, with a small effect size. This is plausible because of the diversity of our subjects. As for the frame time, our results do not allow us to draw any conclusions.

Thus, we also recommend users to use native apps over Web apps for accessing the Internet content when considering their performance. For developers, it may not be realistic to expect that the choice between native and Web apps can be made only based on their expected energy consumption and performance. However, they should not offer a Web app for purposes other than user acquisition. Thus, it may be acceptable to offer only limited functionality in the Web app while committing more to developing a native app and encouraging users to switch to it. Furthermore, researchers may consider the results of this empirical study to motivate their future research on the efficient usage of runtime resources of mobile Web apps and Android browsers.

## VII. Related Work

Oliveira et al. [24] compared the energy efficiency of native Java apps and JavaScript apps based on the Apache Cordova framework (system WebView) running on Android 5 across 33 benchmarks using the same algorithms. In 26 cases, the Java implementation consumed on average 1.82 and 2.09 times energy as the JavaScript version between the two benchmark suites (Rosetta Code and The Computer Language Benchmark Game), but in 5 cases it consumed 1.4 times energy and time. They noted that there seems to be no correlation between energy efficiency and execution time and pointed out that only Java can utilize multiprocessing. JavaScript seems favorable in cases of multiple small computations. Furthermore, the authors hybridized two open-source apps by re-implementing a compute-intensive functionality using different invocation strategies. Their experiment shows that in certain cases this can yield a significant improvement (35.69 times less energy consumed) without seriously impacting the maintainability of the codebase. However, the applicability of this strategy depends on the amount of computation inside the app. They later published a follow-up study, which also includes a comparison to C++ using the Android NDK [25], demonstrating two orders of magnitude performance increases in an application rewritten using a combined approach using Java and C++. The results of our study favour native apps in terms of energy efficiency, which is different from the results obtained by Oliveira et al.; we conjecture that the difference in the results of these two studies lies in the fact that Oliveira et al. focused on the energy consumption of *computation* and did not consider complete apps, including the user interface.

Ma et al. [26] compared native and Web versions of 328 popular services offered by 12 different providers on Android 4.2 and Chrome 40 to investigate if Web apps are generally less performant. Their experiments focused heavily on energy consumption and networking. Ma et al. pointed out that requests made by Web apps are usually larger and that they generally need to fetch more resources compared to native apps, which typically bundle static resources. Even if some resources in Web apps are cached, they expire rather quickly and need to be downloaded again. This might be also an explanation of the results we obtained in this study. Indeed, in the study by Ma et al. Web apps tended to perform more poorly than their native equivalent while consuming less energy. However, in 31% of their experiments, the Web app version provided better performance. The traffic volume was generally lower for Web versions, and caching had the highest impact on GET requests. The difference in utilization of TLS (handshake overhead and cryptography) and the amount of re-fetching per feature utilization between providers and variants caused some significant outliers. While the study by Ma et al. reports energy consumption and detailed Web requests, they do not include hardware utilization and do not group or rank their study subjects, unlike our study which considers the different types of apps listed in Table I based on the content they offer. The Android version used by Ma et al. is also several generations older than the one we use in this study, and might not take advantage of recent optimizations.

Chan-Jong-Chu et al. [27] studied the correlation between performance scores and the energy consumption of 21 out of the 100 most-visited Websites on the Internet, according to the Alexa Rank. Their primary goal was to understand how the performance of Web apps could potentially impact the energy consumption of a mobile device. Their results indicate a significantly lower energy consumption by Web apps that have better performance scores. Our primary goal is to understand the difference in energy consumption and performance between the native and Web version of the same app, while the authors compared different Web apps with regard to their performance and energy consumption.

Corbalan et al. [21] compared the energy efficiency from different frameworks for cross-platform mobile development. The authors classified the frameworks into four different main categories: *Web Approach, Hybrid Approach, Interpreted Approach, and Cross-Compiled Approach*. Their primary goal was to understand which of these frameworks consumed less energy based on certain tasks performed in mobile device experiments. The results show that *Web* and *Hybrid Approaches* consume the most energy, with 3.31% and 3.03% in Android, whilst iOS devices consumed 2.48% and 2.5%. This is in line with the results of our study.

Metri et al. [22] compared apps and Web apps belonging to the categories of browsers, social networking, as well as video and music streaming across Windows 8.1, iOS 7.0.6, and Android 4.3 tablets with respect to energy consumption and hardware utilization. In line with our results, the results obtained by Metri et al. show that native apps usually outperform their Web counterpart in terms of energy efficiency, which they attribute to the higher CPU utilization and lower memory utilization. They also observed on Windows that using multiple cores can speed up certain operations and allow them to sleep for longer periods afterward. Furthermore, they observed the wake-up frequency of the hardware components negatively affects energy efficiency.

## VIII. Threats To Validity

**Internal Validity**. An internal threat to validity may arise from the subject selection. Since we only select 10 subjects from a pool of over 2.6 million as of June 2022 [28], there may be some potential issues of misrepresenting the general population. As a mitigation, we consider app categories, which were previously discussed in Section III, and select exactly 2 apps for each category. This ensures a level of diversity in the subjects.

Another threat would be browser and native app caching, which falls under maturation internal threats since the effect occurs on subsequent runs. Caching could lead to inconsistent results, as the browser or native app may use saved data instead of downloading it on subsequent runs. To mitigate this potential threat, we clear the native app for the corresponding subject or browser cache after every run. Thus, making the app's behavior consistent across runs.

Moreover, apps and their dynamic content may change greatly over time. This means that replications of the experiment may not arrive at the same results and that even the results within the experiment may be influenced by the time of execution. To mitigate this, we provide the exact build numbers of all native apps which are not updated during the course of the study. This is not possible for Web apps and HTTP interfaces. We did not observe a change in the type of dynamic content or update of the Web apps during the experiment.

**External Validity**. First, the experiment is run using only one browser, Google Chrome. Browsers can vary in design and implementation, and these differences could impact the metrics like energy consumption or CPU, and memory utilization measured in this experiment. However, since Google Chrome is the most widely used browser across Android devices (often coming preinstalled) with a market share of over 60% across all mobile operating systems over the last few years [12], no other browser on Android has a comparatively significant market share. Thus, the selection is highly representative.

Second, similar to the browser threat, we are only running this experiment on a single device, as described in Section IV. The device could also be a factor in the experiment, which may influence the outcome and thus be important to consider for the generalizability of the results. Due to the limited scope of the experiment, we did not mitigate this threat. However, it is reasonable to assume that our results are a good indicator for the average contemporary Android based smartphone, as represented by the Nokia 6.2 (model TA-1198) test device. Running Android version 10, which accounts for roughly 20% of all Android phones as of September 2022 [29], the operating system version represents a significant portion of Android devices and can be considered the future lower bound for relevant versions. Future replications of this study might consider different devices under different configurations.

An external threat to validity may arise from the simulated interactions with the app during the experiments. Our experiment is limited to performing a single static loop of basic interaction with each app. Such an approach may fail to represent how the app is used under real circumstances. This threat is not directly mitigated due to the limited scope of the experiment, as well as other practical reasons. It is not possible to include certain activities such as e-commerce checkout and posting content since providers actively prevent this by legal and technical means. These activities would require the use of a 'man in the middle' to completely simulate each remote resource accurately, which is far from trivial. However, such activities are less common than those which only consist of consuming content provided through the app and which can thus be integrated into the experiment. We further took great care in covering the major use cases per app with the scripted interaction to minimize the potential impact of this threat.

**Construct Validity**. To mitigate potential inadequate per-operational explanation of constructs, we defined our constructs a priori before the experiment execution. All the details related to the design of the experiment (*e.g.,* the goal, research questions, variables, data analysis procedures) was defined before executing the experiment. We used the GQM approach to define our goal, which then guided us to derive the research questions of this study. The hypotheses, dependent and independent variables, and treatments were all defined during the planning phase of the experiment.

**Conclusion Validity**. First, since most of the native and Web apps from our selected subjects contain unpredictable interstitial advertisements, this could introduce some 'flaky' behavior when running the experiment. In order to achieve consistent behavior, we had to mitigate this issue by using an external service called dns.adguard.com for blocking advertisements in all native and Web apps based on domain name resolution. For some subjects, this did not remove all advertisements. For example, Pinterest does not use a separate domain to serve advertisements. However, this is ignored, since the advertisements displayed on Pinterest behave similarly to 'normal' posts, so the scripted interaction does not need to handle them separately. System-wide advertisement blocking requires technical awareness and expertise of the user and may thus be rather uncommon, however, it has limited effectiveness and does not influence the behavior of apps in any significant form or their functionality.

Second, variations in device settings, such as the screen brightness or audio playback volume, could influence the measurements for the dependent variables. As a mitigation, we created a custom script that is executed during the experiment setup phase. This script is responsible for handling the state of the device, making sure that all settings are reset as described in detail in Section IV.

Third, another potential threat to the reliability of measurements is the tools used to record the metrics, which are the basis of the dependent variables in our experiment. Due to the limitations of available and compatible profilers, we could not address this issue. The Trepn profiler [30] seemed incompatible with the device used in the experiment. Thus, all measurements are obtained using the internal profiling tools of the Android device as described in Table IV. Before analyzing the data obtained from the measurements, they are scrutinized and all invalid values are removed as described in Section V. This only affected the metric network traffic volume and thus, this metric is not relied on to answer our research questions. Instead, the null hypothesis is rejected on the basis of the mean memory load, which paints a sufficiently clear picture.

Moreover, investigating only 10 pairs of native and Web apps would result in a very small sample with low statistical power. Thus, we perform 25 repetitions, resulting in a total of 500 data points to mitigate this threat. The number of repetitions has been decided based on the literature on measurement-based experiments on mobile (web) apps and available resources.

Due to the limited number of subjects and their potentially large heterogeneity, the dependent variables were not likely to be normally distributed. This results in parametric tests not being applicable. Furthermore, our non-parametric tests may not be adequate to reject the null hypotheses which are concerned with the population mean. This could result in a type 1 error. This is not trivial to address. We use non-parametric tests for non-normally distributed data instead of transforming it. Since the effect size is either very small or very large, we can confidently make a statement regarding the null hypotheses.

## IX. CONCLUSIONS

In this paper we conducted an empirical analysis on the energy consumption and performance of native Android apps and their Web counterparts. In our experiment, we selected 10 Internet content platforms across 5 categories, having 2 subjects per category. Then, we measure the energy consumption, network traffic volume, CPU load, memory load, and frame time of their native and Web versions. Our results show that Web apps consume about 53% more energy than their native counterparts. From our findings, we conclude that Web apps consume significantly more energy than their native counterparts, with a large effect size. In addition, Web apps use more CPU and memory, with a statistically significant difference and large effect size. The CPU utilization of Web apps is roughly 49% higher than their native versions, yet has a slightly lower standard deviation. The difference in memory utilization can most likely be attributed to the overhead posed by the Google Chrome browser. Our results do not allow us to draw any conclusion in terms of the frame time. Based on our findings, we suggest users to use native apps over their Web counterparts for accessing the Internet contents, when possible. Nevertheless, users should also consider other factors when considering to use Web or native apps, such as: available storage space, convenience, availability of the installed native apps, etc.

Possible future work includes extending the experiment by increasing the number of subjects per category to investigate the degree of the observed variance in the measurements based on the app category. Since the use of web browsers may have an impact on energy consumption and performance, it will be interesting to repeat the experiment on other web browsers. Furthermore, it would be interesting to replicate this study on iOS devices as it holds a global market share of 26.98% [31]. Finally, it would be interesting to investigate deeper on the *root causes* of the observed differences between native and Web mobile apps by investigating their source code, system API calls, used programming language (*e.g.,* Kotlin vs Java in the case of native apps [32]), and other technical aspects.

ACKNOWLEDGMENTS

This project has received funding from the European Union's Horizon 2020 research and innovation programme under the Marie Skłodowska-Curie grant agreement No 871342 "uDEVOPS".

REFERENCES

[1] "Desktop vs mobile vs tablet market share worldwide," https://gs.statcounter.com/platform-market-share/desktop-mobile-tablet/worldwide/\#monthly-200901-202209, online; Accessed: December 2022.
[2] "iPhone 13 Specification," https://www.apple.com/iphone-13/specs/, online; Accessed: December 2022.
[3] V. Le Pochat, T. Van Goethem, S. Tajalizadehkhoob, M. Korczyński, and W. Joosen, "Tranco: A research-oriented top sites ranking hardened against manipulation," in *Proceedings of the 26th Annual Network and Distributed System Security Symposium*, ser. NDSS 2019, Feb. 2019.
[4] G. Prakash, "Google Play Store Apps – Google Play Store App data of 2.3 Million+ applications." 2021. [Online]. Available: {https://www.kaggle.com/datasets/gauthamp10/google-playstore-apps}
[5] I. Malavolta, E. M. Grua, C.-Y. Lam, R. de Vries, F. Tan, E. Zielinski, M. Peters, and L. Kaandorp, "A Framework for the Automatic Execution of Measurement-based Experiments on Android Devices," in *35th IEEE/ACM International Conference on Automated Software Engineering Workshops (ASEW '20)*. ACM, 2020. [Online]. Available: https://github.com/S2-group/android-runner/blob/master/documentation/A_Mobile_2020.pdf
[6] "dumpsys," https://developer.android.com/studio/command-line/dumpsys, online; Accessed: December 2022.
[7] O. de Munk, G. L. Scoccia, and I. Malavolta, "The state of the art in measurement-based experiments on the mobile web," *Information and Software Technology*, vol. 149, p. 106944, 2022. [Online]. Available: https://www.sciencedirect.com/science/article/pii/S095058492200091X
[8] J. Cohen, *Statistical Power Analysis for the Behavioral Sciences*. Routledge, july 1988. [Online]. Available: https://doi.org/10.4324/9780203771587
[9] J. Romano, J. Kromrey, J. Coraggio, and J. Skowronek, "Appropriate statistics for ordinal level data: Should we really be using t-test and Cohen'sd for evaluating group differences on the NSSE and other surveys?" in *annual meeting of the Florida Association of Institutional Research*, 2006, pp. 1–3.
[10] "Replication package of this study," https://github.com/S2-group/mobilesoft-2023-app-vs-web-android-rep-pkg, online.
[11] "Android debug bridge (adb)," https://developer.android.com/studio/command-line/adb, online; Accessed: December 2022.

[12] "Mobile browser market share worldwide jan 2009 - sept 2022," https://gs.statcounter.com/browser-market-share/mobile/worldwide/\#monthly-200901-202209, online; Accessed: December 2022.
[13] "Technical specifications - nokia 6.2," https://www.nokia.com/phones/en_int/nokia-6-2/specs, online; Accessed: December 2022.
[14] "Capture a system trace on the command line," https://developer.android.com/topic/performance/tracing/command-line, online; Accessed: December 2022.
[15] "Ookla speed test," https://www.speedtest.net, online; Accessed: December 2022.
[16] Posit, *RStudio: Integrated Development Environment for R*, 2022. [Online]. Available: https://posit.co/products/open-source/rstudio
[17] R Core Team, *R: A Language and Environment for Statistical Computing*, R Foundation for Statistical Computing, Vienna, Austria, 2022. [Online]. Available: https://www.R-project.org/
[18] M. Torchiano, *effsize: Efficient Effect Size Computation*, 2020, r package version 0.8.1. [Online]. Available: https://CRAN.R-project.org/package=effsize
[19] H. Wickham, M. Averick, J. Bryan, W. Chang, L. D. McGowan, R. François, G. Grolemund, A. Hayes, L. Henry, J. Hester, M. Kuhn, T. L. Pedersen, E. Miller, S. M. Bache, K. Müller, J. Ooms, D. Robinson, D. P. Seidel, V. Spinu, K. Takahashi, D. Vaughan, C. Wilke, K. Woo, and H. Yutani, "Welcome to the tidyverse," *Journal of Open Source Software*, vol. 4, no. 43, p. 1686, 2019.
[20] A. Almeida, A. Loy, and H. Hofmann, *ggplot2 Compatible Quantile-Quantile Plots in R*, 2018. [Online]. Available: https://doi.org/10.32614/RJ-2018-051
[21] L. Corbalan, J. Fernandez, A. Cuitiño, L. Delia, G. Cáseres, P. Thomas, and P. Pesado, "Development frameworks for mobile devices: A comparative study about energy consumption," in *Proceedings of the 5th International Conference on Mobile Software Engineering and Systems*, ser. MOBILESoft '18. New York, NY, USA: Association for Computing Machinery, 2018, p. 191–201. [Online]. Available: https://doi.org/10.1145/3197231.3197242
[22] G. Metri, W. Shi, and M. Brockmeyer, "Energy-efficiency comparison of mobile platforms and applications: A quantitative approach," in *Proceedings of the 16th International Workshop on Mobile Computing Systems and Applications*, ser. HotMobile '15. New York, NY, USA: Association for Computing Machinery, 2015, p. 39–44. [Online]. Available: https://doi.org/10.1145/2699343.2699358
[30] "Introducing Trepn Profiler 6.0," https://developer.qualcomm.com/blog/introducing-trepn-profiler-60, online; Accessed: December 2022.
[23] I. A. Qazi, Z. A. Qazi, T. A. Benson, G. Murtaza, E. Latif, A. Manan, and A. Tariq, "Mobile web browsing under memory pressure," *SIGCOMM Comput. Commun. Rev.*, vol. 50, no. 4, p. 35–48, oct 2020. [Online]. Available: https://doi.org/10.1145/3431832.3431837
[24] W. Oliveira, W. Torres, F. Castor, and B. H. Ximenes, "Native or web? a preliminary study on the energy consumption of android development models," in *2016 IEEE 23rd International Conference on Software Analysis, Evolution, and Reengineering (SANER)*, vol. 1, 2016, pp. 589–593.
[25] W. Oliveira, R. Oliveira, and F. Castor, "A study on the energy consumption of android app development approaches," in *2017 IEEE/ACM 14th International Conference on Mining Software Repositories (MSR)*, 2017, pp. 42–52.
[26] Y. Ma, X. Liu, Y. Liu, Y. Liu, and G. Huang, "A tale of two fashions: An empirical study on the performance of native apps and web apps on android," *IEEE Transactions on Mobile Computing*, vol. 17, no. 5, pp. 990–1003, 2018.
[27] K. Chan-Jong-Chu, T. Islam, M. M. Exposito, S. Sheombar, C. Valladares, O. Philippot, E. M. Grua, and I. Malavolta, "Investigating the correlation between performance scores and energy consumption of mobile web apps," in *Proceedings of the Evaluation and Assessment in Software Engineering*, ser. EASE '20. New York, NY, USA: Association for Computing Machinery, 2020, p. 190–199. [Online]. Available: https://doi.org/10.1145/3383219.3383239
[28] "Number of available applications in the google play store from december 2009 to september 2022," https://www.statista.com/statistics/266210/number-of-available-applications-in-the-google-play-store/, online; Accessed: December 2022.
[29] "Mobile android version market share worldwide sept 2018 - sept 2022," https://gs.statcounter.com/android-version-market-share/mobile/worldwide/\#monthly-201809-202209, online; Accessed: December 2022.
[31] "Mobile Operating System Market Share Worldwide," https://gs.statcounter.com/os-market-share/mobile/worldwide, online; Accessed: December 2022.
[32] M. Peters, G. L. Scoccia, and I. Malavolta, "How does migrating to kotlin impact the run-time efficiency of android apps?" in *21st IEEE International Working Conference on Source Code Analysis and Manipulation (SCAM)*, September 2021, pp. 36–46. [Online]. Available: http://www.ivanomalavolta.com/files/papers/SCAM_2021.pdf